\documentclass[reprint,aps,notitlepage]{article}
\usepackage{authblk}
\usepackage[numbers,sort&compress]{natbib}

\usepackage{amsmath,amssymb}
\usepackage{soul}
\usepackage[normalem]{ulem}
\usepackage{hyperref}
\usepackage{xcolor}
\usepackage{graphicx}
\usepackage{upgreek}
\usepackage{bm} 
\usepackage{dsfont}

\newcommand{\xref}[1]{(\ref{#1})}
\newcommand{\bs}[1]{\bm{#1}}
\newcommand{\mbf}[1]{\mathbf{#1}}
\usepackage{dsfont} 

\usepackage{diagbox}

\usepackage{supertabular}

\newcommand{\ra}{\rangle}
\newcommand{\la}{\langle}
\newcommand{\hx}{\hat{x}}
\newcommand{\hp}{\hat{p}}
\newcommand{\hrho}{\hat{\rho}}

\newcommand{\ih}{i\hbar}

\newcommand{\eg}{e.g., }
\newcommand{\cf}{cf.\ }
\newcommand{\be}{\begin{equation}}
\newcommand{\ee}{\end{equation}}
\newcommand{\bp}{\mbf{p}}
\newcommand{\bx}{\mbf{x}}
\newcommand{\brho}{\bs{\rho}}
\newcommand{\bxi}{\bs{\xi}}

\usepackage{soul}

\begin{document}

\title{Remarks on the quasi-position representation in models of generalized uncertainty principle}

\author[]{Andr\'e Herkenhoff Gomes\thanks{andre.gomes@ufop.edu.br}}
\affil[]{\small Departamento de F\'isica, Universidade Federal de Ouro Preto,\\ Ouro Preto, MG, Brazil}%

\date{}

\maketitle

\begin{abstract}
This note aims to elucidate certain aspects of the quasi-position representation frequently used in the investigation of one-dimensional models based on the generalized uncertainty principle (GUP). We specifically focus on two key points: (i) Contrary to recent claims, the quasi-position operator can possess physical significance even though it is non-Hermitian, and (ii) in the quasi-position representation, operators associated with the position, such as the potential energy, also behave  as a derivative operator on the quasi-position coordinate, unless the method of computing expectation values is modified. The development of both points revolves around the observation that the position and quasi-position operators share the same set of eigenvalues and are connected through a non-unitary canonical transformation. This outcome may have implications for widely referenced constraints on GUP parameters.
\end{abstract}

\section{Introduction}
\label{sec:introduction}

Quantum gravity phenomenology at low energies often involves the consideration of corrections to the Heisenberg uncertainty principle. These corrections arise when gravity plays a significant role in quantum phenomena, leading to the so-called generalized uncertainty principles (GUPs) \cite{mead,maggiore,scardigli} --- see also, \eg \cite{garay,hossen2013,tawfik-diab-2014,tawfik-diab-2015,igup,bosso-luciano-petruzziello-wagner} for reviews and experimental constraints.

In this note, our focus is on one-dimensional models as a theoretical framework for studying GUPs of the form
\be\label{gup}
\Delta x \Delta p \ge \frac{\hbar}{2} f(\Delta p),
\ee
where $f$ is a model-dependent real function that reduces to unity whenever quantum gravity is neglected or when $\la \bp^2\ra$ tends to zero. Previous studies have utilized heuristic approaches to explore the effects of GUPs on gravity in various contexts, ranging from astrophysics to cosmology, leading to valuable insights on quantum gravity \cite{cosmo1,scardigli-casadio,grav1,grav2,stars,casadio-scardigli-2020,cosmo2,grav3}.

In addition to the heuristic approaches, an algebraic approach to GUPs has been widely used in the investigation of quantum mechanical systems \cite{maggiore-algebra,kmm95}. This approach sets the commutation relation between position and momentum operators, given by\footnote{In this note, operators are written in boldface, eigenvalues in italic, and operators in specific representation in italic letters with a hat --- \eg $\bx$ is the position operator with eigenvalues $x$ and denoted by $\hx$ in momentum space, thus $\hx\la p|x\ra \equiv \la p|\bx|x\ra = x\la p|x\ra$.}
\be\label{comm}
[\bx,\bp] = \ih f(\bp),
\ee
as fundamental since it defines the kinematics of the model.\footnote{It should be noted that \xref{comm} implies \xref{gup} through the Robertson-Schr\"odinger relation $\Delta x \Delta p \ge \frac{1}{2}|\la[\bx,\bp]\ra|$ \cite{robertson,schrodinger} only when the inequality $\la f(\bp) \ra \ge f(\Delta p)$ is satisfied for physical states. However, this condition holds only for specific cases and not for all models --- \eg it holds for $f(\bp) = 1+\beta \bp^{2}$, but not for $f(\bp) = \sqrt{1 + \beta \bp^2}$. Therefore, the commutation relation \xref{comm}, rather than the GUP \xref{gup}, is generally considered more fundamental in the algebraic approach.} An advantage of this approach is that the standard machinery of quantum mechanics can still be used once it is found a representation for $\bx$ and $\bp$ satisfying the above commutator. However, models based on a GUP often feature a nonvanishing lower bound on the uncertainty of position measurements \cite{mangano2016,bosso-petruzziello-wagner},
\be\label{minimum}
\la \psi | (\Delta \bx)^2 |\psi\ra \ge (\Delta x)^2_\text{min} > 0,
\ee
and this adds an extra layer of technical challenge as it prevents using the wave function representation in position space \cite{kmm95}. One way of amending this and recovering information on position is to approach quantum mechanics in the so-called quasi-position representation, which hinges on states $|\xi\ra$ of maximal localization (ML) around the position $\xi$ under the minimal position uncertainty constraint \cite{kmm95,detournay,bernardo-esguerra-maximally2,pasquale-xbasis}:
\be\label{ml-definition}
\la\xi|\bx|\xi\ra \equiv \xi,
\qquad
\la\xi|(\Delta \bx)^2|\xi\ra \equiv (\Delta x)^2_\text{min}.
\ee
Many investigations have been done within this approach, including studies of the square potential well \cite{nozari2006,ali-das-vagenas09,pedram-europhys,pedram2010,ali-das-vagenas11,pedram12-plb3,pedram12-prd,blado,bernardo-esguerra,Shababi16,Shababi18,chung-hass2019,nogueira2020,Shababi20,pasquale-xbasis}, quantum scattering \cite{scattering}, toy models of quantum field theory \cite{bosso-luciano}, time evolution by means of Euclidean path-integral \cite{bernardo-esguerra-maximally1,bernardo-esguerra-maximally2}, noncommutative quantum theories \cite{lubo}, and the Casimir effect \cite{casimir1,casimir2,casimir3}.

It is important to note that the quasi-position representation does not encode the quasi-position $\xi$ as the eigenvalue of an observable associated with a quasi-position operator $\bxi$. Instead, $\xi$ is solely considered as an average position and the ML state $|\xi\ra$ is viewed solely as a state that satisfies \xref{ml-definition}. In contrast to previous arguments presented in \cite{pasquale-xbasis}, which suggest that $\bxi$ cannot be interpreted as a position coordinate due to its non-Hermitian nature, in this note we propose that $\bxi$ can indeed be regarded as a position operator. We demonstrate that $\bx$ and $\bxi$ are related by a non-unitary canonical transformation and they share the same set of eigenvalues. The non-Hermitian nature of $\bxi$ is not an obstacle but rather reflects the inherent fuzziness expected in position measurements. According to this interpretation, the observable relevant in the context of quantum gravity phenomenology may be the quasi-position operator, rather than $\bx$ itself. However, we show that when calculating expectation values and expressing the Hamiltonian in terms of $\bxi$ (or simply $\xi$ in quasi-position representation), an additional level of caution is necessary. In particular, the potential energy, when expressed in the quasi-position representation, also acts as a derivative operator on the quasi-position coordinate, unless the method of computing expectation values is adjusted.

To the best of our knowledge, this aspect has not been addressed in the existing literature and has the potential to impact frequently cited bounds on GUP parameters. Moreover, the relevance of this work goes beyond the realm of quantum gravity, having applications in ordinary quantum mechanics as well, particularly in condensed matter systems that involve the concept of an effective minimal position uncertainty, such as analogue models of quantum gravity.

This note is organized as follows. In the next section, we present an overview of operator representations in the one-dimensional GUP models to introduce relevant results that are referred to in Section \ref{sec:operator}, which focuses on the study of the quasi-position operator. At last, in Section \ref{sec:concluding-remarks}, we examine the implications on the representation of Hamiltonians and conclude with our final remarks.

\section{Overview of Operator Representations}
\label{sec:overview}

\subsection{Momentum representations}

In the literature, there are two commonly used representations related to momentum. One is based on the eigenvectors $|p\ra$ of the momentum operator $\bp$ itself. Physical states are denoted as $\psi(p)\equiv \la p|\psi\ra$, and the following representation satisfies the commutator \xref{comm}:
\be\label{p-rep}
\bx|\psi\ra \to \ih f \partial_p \psi(p),
\qquad
\bp|\psi\ra \to p \psi(p),
\ee
where $\partial_p \equiv d/dp$ for simplicity. To ensure that the position operator $\bx$ is a symmetric operator, $\la\psi|\bx|\phi\ra = \la\phi|\bx|\psi\ra^*$, the momentum eigenvectors satisfy the resolution of identity and orthonormality relation as follows:
\be
\mathds{1} = \int_{-\infty}^\infty \frac{dp}{f(p)} |p\ra\la p|,
\qquad
\la p|p'\ra = f(p)\delta(p-p').
\ee
Although this is the simplest representation in momentum space, it's worth noting that other representations can be obtained through a non-unitary canonical transformation given by $\hx \to \hx + \frac{1}{2}\ih\varepsilon \partial_p f$, where $\hx \equiv \ih f\partial_p$ and $\varepsilon$ is a real arbitrary parameter. This transformation also induces a change in the integration measure, $dp/f \to dp/f^{1-\varepsilon}$, which keeps $\bx$ symmetric and preserves expectation values \cite{andre-canonical}.

The other representation is based on eigenvectors $|\rho\ra$ of the generator of spatial translations $\brho$. This generator is defined as the operator that is canonically conjugate to $\bx$, satisfying the commutation relation  $[\bx,\brho]=\ih$.\footnote{The operator $\brho$ is sometimes referred to as the auxiliary or unmodified momentum operator (\eg \cite{pasquale-xbasis,mangano2016}, respectively). Alternatively, it is indirectly associated with a wave number operator $\brho=\hbar\mbf{k}$ (\eg \cite{bosso-petruzziello-wagner}). However, we choose to use the term ``generator of translations'' to describe it, as originally presented in \cite{kempf-mangano97}.} In this basis,
\be
\bx|\psi\ra \to \ih \partial_\rho \psi(\rho),
\qquad
\bp|\psi\ra \to p(\rho)\psi(\rho),
\ee
where $\psi(\rho) \equiv \la\rho|\psi\ra$ and with $p(\rho)$ such that
\be\label{rho-p}
\frac{dp(\rho)}{d\rho} = f(p(\rho))
\qquad \text{or} \qquad
\rho(p) = \int_0^p \frac{dp'}{f(p')}.
\ee
It should be noted that only in the limit $f\to1$ do $p$ and $\rho$ become equivalent. For $f\neq1$, the requirement of invertibility for $\rho(p)$ restricts it to be a monotonic function possibly defined only in a restricted interval, which we assume here to be symmetric for definiteness. Consequently, $f(p)$ does not change sign and, specifically, it is not an odd function \cite{pasquale-xbasis}. A useful result derived from this representation is \cite{mangano2016,bosso-petruzziello-wagner}
\be\label{delta-x-rho}
(\Delta x)_\text{min} = 
 \frac{\pi\hbar}{2\rho_\text{max}}.
\ee

\subsection{Position representations}

Before discussing the quasi-position representation, a brief remark on the position space representation is in place. In ordinary quantum mechanics, a state $|\psi\ra$ belonging to the Hilbert space $\mathcal{H}$ represents a physical state for a system if it satisfies the corresponding Schr\"ondiger equation and relevant boundary conditions. Once a representation for $\bx$ and $\bp$ is selected to fulfill $[\bx,\bp]=\ih$, it incorporates the Heisenberg uncertainty principle into the Schrödinger equation and wave function $\psi$. Position eigenvectors $|x\ra$ are not physical states \cite{cohen1}, but they can be approached by physical states of increasingly precise localization in position space since $(\Delta x)_\text{min}=0$, \eg $|\psi\ra = \int dx \psi(x) |x\ra \to |y\ra$ as $\psi(x) \equiv \la x|\psi\ra \to \delta(x-y)$, which requires $\bx$ to have real eigenvalues and orthonormal eigenvectors (self-adjoint operator). However, this conclusion does not hold in the context of GUPs if $(\Delta x)_\text{min} > 0$. In such cases, position eigenvectors can no longer be arbitrarily approached by physical states \cite{kmm95}. Instead, physical states belong to a subset $\mathcal{S} \supset \mathcal{H}$ that is disjoint from the subset containing position eigenvectors. Thus, the representation
\be\label{forbidden}
|\psi\ra \to \psi(x) = \la x | \psi \ra
\qquad
\text{(unphysical)}
\ee
is generally unphysical when there exists a non-zero minimum uncertainty in position. Consequently, the position space representation
\be\label{alsoforbidden}
\bx|\psi\ra \to x \psi(x),
\qquad
\brho|\psi\ra \to -\ih \partial_x \psi(x),
\qquad
\text{(unphysical)}
\ee
is also unphysical even though it satisfies $[\bx,\brho]=\ih$. Functional analysis of $\bx$ reveals it is symmetry, but not self-adjoint, and still have real eigenvalues, though now associated to non-orthonormal eigenvectors \cite{kempf2000,pedram12-plb2,nozari-etemadi}.

Information about position can be obtained through position-based representation using physical states $|\xi\ra$ of maximal localization (ML) \cite{kmm95,bernardo-esguerra-maximally2,detournay,pasquale-xbasis}. These states minimize the uncertainty in position around a specific point  $x=\xi$, while adhering to the constraint of minimal position uncertainty --- see defintion \xref{ml-definition}. The explicit form of the unnormalized ML state in momentum space is given by a modified plane wave \cite{pasquale-xbasis}:
\be\label{planewave}
\la p|\xi \ra = A(p) e^{-i\xi\rho(p)/\hbar}.
\ee
Here, $\rho=\rho(p)$ is determined by equation \xref{rho-p} and the momentum-dependent amplitude $A(p)$ is a real function defined as
\be\label{amplitude}
A(p) 
= \text{exp}\left\{
-\frac{\la f(\bp) \ra}{2(\Delta p)^2} \left( \int dp \frac{p-\la \bp \ra}{f(p)} \right)
\right\},
\ee
where $\la f(\bp)\ra$, $\Delta p$, and $\la\bp\ra$ are quantities with predetermined values that ensure the minimum uncertainty in position in the ML state $|\xi\ra$. From this, a generalized Fourier transformation relating $\psi(p)\equiv\la p|\psi\ra$ and $\psi(\xi)\equiv\la \xi|\psi\ra$ is defined as \cite{kmm95,pasquale-xbasis} 
\be\label{fourier}
\psi(\xi) = \int_{-\infty}^\infty \frac{dp}{f(p)} \la\xi|p\ra \psi(p),
\qquad
\psi(p) = \frac{1}{2\pi\hbar} \int_{-\infty}^\infty d\xi \la\xi|p\ra^{-1} \psi(\xi).
\ee
By performing straightforward calculations, the quasi-position representations of $\bx$ and $\bp$ is finally found to be
\be\label{xi-rep}
\bx|\psi\ra \to
\left[
\xi + \ih \frac{\la f(\hp) \ra}{2(\Delta p)^2} \left( \hat{p} - \la \hp \ra \right)
\right]\psi(\xi),
\qquad
\bp|\psi\ra \to \hat{p}\psi(\xi),
\ee
where we denote $\hat{p} \equiv p(\hrho)$ to make it explicit that $p$ is a function of the differential operator $\hrho=-\ih\partial_\xi$ in this representation. It is worth noting that, as a consequence, the potential energy behaves as a derivative operator as well,
\be
V(\bx) |\psi\ra \to V(\xi,\partial_\xi)\psi(\xi).
\ee
In particular, disregarding the derivative term in the quasi-position representation of $\bx$ generally leads to first-order errors in the GUP parameters.

\section{The quasi-position operator}
\label{sec:operator}

In the Introduction, we advocate for the physical status of an operator $\bxi$ that satisfies $\bxi|\xi\ra=\xi|\xi\ra$. This quasi-position operator can be guessed by inspection of equation  \xref{xi-rep},
\be\label{xi-operator}
\bxi \equiv \bx + \ih \frac{\la f(\bp)\ra}{2(\Delta p)^2} (\bp - \la\bp\ra).
\ee
The appearent sign difference between \xref{xi-rep} and \xref{xi-operator} is justified by noticing that the term $\xi \psi(\xi)$ in \xref{xi-rep} arises from $\la\xi|\bxi^\dagger|\psi\ra$, not $\la\xi|\bxi|\psi\ra$ --- however, it is important to note that $\bx$ is a symmetric operator and can be expressed equivalently using either $\bxi$ or $\bxi^\dagger$. The quasi-position operator is clearly non-Hermitian or, more precisely, not symmetric, but we see next that its eigenvalues are real and coincide with those of $\bx$. This indicates that $\xi$ represents the actual position eigenvalue in a ML state, rather than just an average position as suggested by the definition \xref{ml-definition}. Even though $\Delta \xi = 0$ in a ML state, the non-Hermiticity of $\bxi$ manifests itself in the non-orthogonality of ML states and realize the fuzziness of position measurements since $\Delta x = (\Delta x)_\text{min}$ still holds for such states.

\subsection{Canonical transformation and uncertainty relation}

We propose that the quasi-position operator $\bxi$ can be defined as the outcome of a non-unitary, commutator preserving transformation on $\bx$ implemented by the Hermitian operator $A(\bp)$; namely,
\be\label{appendix-transf}
\bxi \equiv \bx' = A(\bp) \bx A^{-1}(\bp),
\qquad
A(\bp) 
\equiv \text{exp}\left\{
-\frac{\la f(\bp) \ra}{2(\Delta p)^2} \left( \int d\bp \frac{\bp-\la \bp \ra}{f(\bp)} \right)
\right\},
\ee
with the expectation values in $\la f(\bp)\ra$, $\Delta p$, and $\la\bp\ra$ taken in the ML state $|\xi\ra$ and assumed to imply minimum $\Delta x$.\footnote{In ordinary quantum mechanics, $(\Delta x)_\text{min} \to 0$ for $\Delta p \to \infty$ so that $\bxi\to\bx$, as expected.} This transformation reproduces the definition of $\bxi$ as shown in \xref{xi-operator} indeed:
\be\label{appendix-definition}
\bxi = \bx + A(\bp)[\bx,A^{-1}(\bp)] = \bx + \ih \frac{\la f(\bp)\ra}{2(\Delta p)^2} (\bp - \la\bp\ra).
\ee
The last equality can be easily confirmed, particularly when working in momentum representation using $[\hx,A^{-1}(p)]=\ih f \partial_p A^{-1}$. We observe that the eigenvalue equation $\bxi|\xi\ra = \xi|\xi\ra$ can be projected onto momentum eigenvectors, leading to the differential equation\footnote{\cf equation (3) in \cite{pasquale-xbasis}.}
\be
\la p| \left[
\bx + \ih \frac{\la f(\bp)\ra}{2(\Delta p)^2} (\bp - \la\bp\ra)
\right]|\xi\ra
= \left[
\ih f\partial_p + \ih \frac{\la f(\bp)\ra}{2(\Delta p)^2} (p - \la\bp\ra)
\right] \la p|\xi\ra
= \xi \la p|\xi\ra.
\ee
The solution $\la p|\xi\ra$ is the momentum space representation  of the ML state given in equation \xref{planewave}, as expected.

The quasi-position operator $\bxi$ differs from the usual position operator in an important aspect: it is not Hermitian, which means that its properties are somewhat different from what one might expect. In particular, the relationship between the operator and its Hermitian conjugate $\bxi^\dagger$ is
\be\label{xi-dagger}
\bxi^\dagger = \bxi - \ih \frac{\la f(\bp)\ra}{(\Delta p)^2} (\bp - \la\bp\ra),
\ee
from which follows that $\la\bxi^\dagger\ra = \la\bxi\ra$ generally holds only in ML states. Despite this difference, the commutator algebra between position and momentum is preserved: $[\bx,\bp]=[\bxi,\bp]=[\bxi^\dagger,\bp]=\ih f(\bp)$. This means that measurements of $\bx$ are still subjected to the mininum uncertainty constraint \xref{minimum}. However, it does not mean that an analogous constraint holds for $\bxi$. Instead, derivation of the uncertainty principle for $\bxi$ and $\bp$ in a ML state gives
\be\label{appendix-xi-p}
\Delta\xi\Delta p \ge 0,
\ee
which is trivially satisfied as an equality since $\Delta\xi = 0$ in ML states.

The proof of \xref{appendix-xi-p} follows the textbook derivation of the Heisenberg uncertainty principle of ordinary quantum mechanics, but for the non-Hermitian operator $\bxi$ instead of $\bx$. It starts from the positivity of the following norm:
\be
\left| \left| \left[ \bxi - \la\bxi\ra + \frac{\la[\bxi,\bp]\ra}{2(\Delta p)^2} (\bp - \la\bp\ra) \right] |\xi\ra \right| \right|^2
=
\left| \left| \left[ \Delta\bxi + \ih\frac{\la f(\bp)\ra}{2(\Delta p)^2} \Delta\bp \right] |\xi\ra \right| \right|^2 \ge 0,
\ee
with all expectation values again taken in the ML state. Defining $(\Delta\xi)^2 \equiv ||\Delta\bxi|\xi\ra||^2 = \la \Delta\bxi^\dagger\Delta\bxi \ra $, the above reads
\be
(\Delta\xi)^2 + \ih\frac{\la f(\bp)\ra}{2(\Delta p)^2} \la \bxi^\dagger\bp - \bp\bxi \ra + \left( \frac{\hbar \la f(\bp)\ra}{2(\Delta p)^2} \right)^2 (\Delta p)^2 \ge 0,
\ee
where the second term above appears after computing $\la\Delta\bxi^\dagger \Delta\bp - \Delta\bp \Delta\bxi\ra = \la \bxi^\dagger\bp - \bp\bxi \ra$. If $\bxi$ were Hermitian, the above would result in the GUP \xref{gup} with $\Delta\xi$ instead of $\Delta x$, but this is not the case. By using the relation \xref{xi-dagger}, we find that $\la \bxi^\dagger\bp - \bp\bxi \ra=0$. This leads us to the inequality
\be
(\Delta\xi)^2 (\Delta p)^2 + \frac{\hbar^2}{4} |\la f(\bp)\ra|^2 \ge 0,
\ee
and ultimately to the uncertainty relation in \xref{appendix-xi-p}. On the one hand, the formal Hermitian nature of $\bx$ is crucial in leading to the GUP; on the other, the non-Hermiticity of $\bxi$ leads to an appearent classical uncertainty relation when the system is in a ML state --- but notice that, by definition, ML states have a built-in minimal uncertainty in position.

\subsection{Eigenvalues and expectation values}

When we carry out a canonical transformation on an operator, it not only changes the operator itself but also affects the basis vectors. In this case, the eigenstates of $\bx$ and $\bxi$ are related by\footnote{This relationship implies $\la p|\xi\ra = A(p) \la p|x\ra$ and also agrees with the momentum representation of the ML state \xref{planewave}. To wit, $\la p|x\ra = e^{-i\xi\rho(p)/\hbar}$ is the position eigenvector in momentum space, obtained from $\la p|\bx|x\ra = \ih f(p)\partial_p \la p|x\ra = \xi\la p|x\ra$.}
\be\label{appendix-basis}
|\xi\ra = A(\bp) |x\ra.
\ee
It is an immediate consequence that both operators share the same set of eigenvalues $\xi \equiv x$:
\be
\bxi |\xi\ra = A(\bp)\bx|x\ra = A(\bp) \xi |x\ra = \xi A(\bp) |x\ra = \xi|\xi\ra,
\ee
where we used $[A(\bp),\xi] = 0$, since $\xi$ represents a number. However, the expectation values for $\bx$ and $\bxi$ should be compared with care as they are generally different. In more detail, the basis change \xref{appendix-basis} also changes general state vectors,
\be
|\psi'\ra = A(\bp) |\psi\ra.
\ee
Thus,
\be\label{expectation-value}
\la \psi|\bx|\phi\ra = \la \psi'|A^{-2}(\bp)\bxi|\phi'\ra,
\ee
meaning that expectation values are preserved provided a change in integration measure, \eg from $[dp/f(p)]$ to $[dp/f(p)] \times A^{-2}(p)$ in momentum space. 

A consequence is that $\la\psi|\bx|\psi\ra$ is generally not equal to $\la\psi|\bxi|\psi\ra$ since $\la\psi|\bp|\psi\ra - \la \bp \ra$ is not equal to zero in \xref{xi-operator}, except in ML states $|\psi\ra = |\xi\ra$, where equality is reached,
\be\label{expected-xi}
\la\xi|\bxi|\xi\ra = \la\xi|\bx|\xi\ra = \xi.
\ee
The last equality coincides with the first of the two conditions in \xref{ml-definition} characterizing the ML state indeed. Making use of \xref{xi-operator}, we also find
\be\label{expected-xi2}
\la\xi|\bxi^\dagger \bxi|\xi\ra
= 
\xi^2
= \la\xi|\bx^2|\xi\ra - \frac{\hbar^2}{4} \frac{\la f(\bp)\ra^2}{(\Delta p)^2}.
\ee
From \xref{expected-xi} and \xref{expected-xi2}, the second condition defining ML states in \xref{ml-definition} is also verified:
\be
\la\xi| (\Delta \bx)^2 |\xi\ra = \la\xi|\bx^2|\xi\ra - \la\xi|\bx|\xi\ra^2 = \frac{\hbar^2}{4} \frac{\la f(\bp)\ra^2}{(\Delta p)^2} = (\Delta x)^2_\text{min},
\ee
where the last equality is a valid identification once it is taken into account the GUP \xref{gup} and the definition in \xref{appendix-transf} that ML states are such that $\la f(\bp)\ra$ and $\Delta p$ above are such that the position uncertainty is minimal.

\subsection{Resolution of identity}

At last, we verify that quasi-position eigenvectors are not mutually orthogonal but can still be used to express the identity operator $\mathds{1}$ as a sum of projections onto ML states. The non-orthogonality becomes evident when we notice that the overlap between ML states, given by
\be
\la\xi'|\xi\ra
= \la\xi'|\left[ \int_{-\infty}^\infty \frac{dp}{f(p)}|p\ra\la p| \right] |\xi\ra
= \int_{-\infty}^\infty \frac{dp}{f(p)} A^2(p) e^{-i(\xi-\xi')\rho(p)/\hbar},
\ee
is not equal to $\delta(\xi'-\xi)$, unless $f(p)=1$ and $\Delta p \to \infty$ ($A\to1$), which restores ordinary quantum mechanics. This arises from the finite resolution for position measurements resulting from $f\neq1$ in the GUP \xref{gup}, or more explicitly, from the non-orthogonality of position eigenstates. This non-orthogonality can be recovered from the above in the limit $\Delta p \to \infty$, where $A\to1$ and $|\xi\ra\to|x\ra$:
\begin{align}
\la x'|x\ra
& = \frac{1}{2\rho_{\text{max}}} \int_{-\infty}^\infty \frac{dp}{f(p)} e^{-i(\xi-\xi')\rho(p)/\hbar}
\nonumber\\[8pt]
& = \frac{1}{2\rho_{\text{max}}} \int_{-\rho_\text{max}}^{\rho_\text{max}} d\rho e^{-i(\xi-\xi')\rho(p)/\hbar} \nonumber\\[8pt]
& = \frac{2(\Delta x)_\text{min}}{\pi(\xi-\xi')} \sin\left[\frac{\pi(\xi-\xi')}{2(\Delta x)_\text{min}}\right].
\end{align}
Here, the normalization factor $1/2\rho_\text{max}$ is introduced to ensure $\la x|x\ra=1$ and the result is expressed in terms of $(\Delta x)_\text{min}$ by means of the relation \xref{delta-x-rho}.

To achieve resolution of the identity in terms of ML states, we propose
\be
\mathds{1} = \frac{1}{2\pi\hbar} \int_{-\infty}^\infty d\xi A^{-2}(\bp) |\xi\ra\la \xi|. \ee
This type of completeness relation, involving ML states, has appeared previously in studies of models with $f(\bp)=1+\beta\bp^2$ within the context of a quantum theory of free scalar fields \cite{matsuo-shibusa}, as well as in the formulation of the Euclidean path integral formalism \cite{bernardo-esguerra-maximally1,bernardo-esguerra-maximally2}. The above relation agrees with the specific relation described in these references and generalizes it for any reasonable $f(\bp)$. The consistency of this proposal can be verified by noticing that it reproduces the Fourier transform given by \xref{fourier},
\begin{align}
\psi(p)
& = \la p| \left( \frac{1}{2\pi\hbar}
\int_{-\infty}^\infty d\xi A^{-2}(\bp) |\xi\ra\la \xi| \right) |\psi\ra
\nonumber\\[8pt]
& = \frac{1}{2\pi\hbar} \int_{-\infty}^\infty \frac{d\xi}{A^2(p)} \la p|\xi\ra \psi(\xi)
\nonumber\\[8pt]
& = \frac{1}{2\pi\hbar} \int_{-\infty}^\infty d\xi \la\xi|p\ra^{-1} \psi(\xi),
\end{align}
and the orthonormality of momentum eigenvectors,
\begin{align}
\la p|p'\ra
& = \frac{1}{2\pi\hbar} \int_{-\infty}^\infty \frac{d\xi}{A^2(p)} \la p|\xi\ra \la\xi|p'\ra
\nonumber\\[8pt]
& = \frac{1}{2\pi\hbar} \int_{-\infty}^\infty d\xi e^{-i\xi[\rho(p)-\rho(p')]/\hbar}
\nonumber\\[8pt]
& = \delta[\rho(p)-\rho(p')] = f(p)\delta(p-p').
\end{align}

\section{Discussion and concludings remarks}
\label{sec:concluding-remarks}

The physical significance of the position $\bx$ and quasi-position $\bxi$ operators is found to be equivalent in the restricted sense that:
\begin{enumerate}
    \item Both share the same set of eigenvalues: $x=\xi$.
    \item Both have the same fundamental commutation relation with the momentum operator: $[\bx,\bp]=[\bxi,\bp]=\ih f(\bp)$.
    \item Measurements of $\bxi$ in maximal localization states yield the minimal position uncertainty: $\Delta x = (\Delta x)_\text{min}$.
    \item Expectation values of the position and quasi-position operators are related by \xref{expectation-value}: $\la \psi|\bx|\phi\ra = \la \psi'|A^{-2}(\bp)\bxi|\phi'\ra$.
\end{enumerate}
Considering this, we recall an idea put forward in \cite{bernardo-esguerra-maximally2}: to assume that the quasi-position $\xi$ serves as the dynamical variable typically associated with position, and to express the Hamiltonian in terms of $\bxi$ and $\bp$. However, we note that $H(\bx,\bp) \neq H(\bxi,\bp)$. Therefore, to effectively implement this idea while remaining consistent with the four points above, it is necessary to appropriately transform the Hamiltonian:
\be
H(\bx,\bp) = A(\bp)^{-1}H(\bxi,\bp)A(\bp).
\ee
Consequently, the time-independent Schrödinger equation transforms to $H(\bxi,\bp)|\psi'\ra = E|\psi'\ra$ or, in quasi-position representation,
\be
H(\xi,\hp)\psi'(\xi) = E \psi'(\xi).
\ee
As before, $\hp=p(\hrho)$ and $\hrho=-\ih\partial_\xi$, and the relation between $\hp$ and $\hrho$ is found from \xref{rho-p}. One appealing aspect is that the quasi-position representation of $\bx$ used here does not require taking derivatives with respect to $\xi$, unlike the approach described in equation \eqref{xi-rep}.

This approach distinguishes itself from the unphysical position representation described in \xref{alsoforbidden} by computing expectation values of position-dependent operators differently. Instead of simply evaluating $\la \psi|\bxi|\phi\ra$, expectation values are computed as $\la \psi|\bx|\phi\ra = \la \psi'|A^{-2}(\bp)\bxi|\phi'\ra$. This distinction is crucial, particularly when performing perturbation theory calculations in the quasi-position representation.

Far from being trivial, the above provides a consistent approach to quantum mechanics in the quasi-position representation. Additionally, it clarifies the seemingly confused literature on  GUP in relation to this topic. It is not uncommon to come across studies where the Schrödinger equation is expressed in the unphysical position representation \xref{alsoforbidden}. Benefiting from hindsight, one might wish to justify this unphysical approach by reinterpreting $x$ as the quasi-position $\xi$. However, one should distinguish between two equivalent possibilities for representing the potential energy:
\be
V(\bx) |\psi\ra \to V(\xi,\partial_\xi) \psi(\xi),
\qquad \text{and} \qquad
V(\bx) |\psi\ra \to V(\xi) \psi'(\xi).
\ee
The first involves the wave function $\psi(\xi)=\la\xi|\psi\ra$, which represents the Fourier transform of the momentum space wave function $\psi(p)$ to quasi-position space, and has the potential $V$ depending also on derivatives [\cf \xref{xi-rep}], which can be inconvenient. The second involves the wave function $\psi'(\xi)=\la\xi|A(\bp)|\psi\ra$ obtained by transforming $\bx$ to $\bxi$ and has $V$ with a simple dependence on $\xi$. To highlight the significance of this conclusion, we notice that the commonly used expressions in the literature,
\be
V(\bx) |\psi\ra \to V(x) \psi(x)
\qquad \text{and} \qquad
V(\bx) |\psi\ra \to V(\xi) \psi(\xi),
\qquad \text{(incorrect)}
\ee
do not generally correspond to the correct choices.

Overlooking this conclusion can potentially lead to the introduction of first-order errors in the GUP parameters when performing calculations in the quasi-position representation. In particular, we have observed this occurrence in the derivation of commonly cited bounds on the GUP parameters obtained from quantum mechanical systems (\eg \cite{das-vagenas08,ali-das-vagenas11,brau}).\footnote{Recent summaries of bounds on GUP parameters can be found, \eg in \cite{bosso-luciano-petruzziello-wagner,igup}.} In the end, we advise approaching these bounds with caution or reexamining them.

\bibliographystyle{apsrev4-1}
\bibliography{references}

\begin{thebibliography}{56}%
\makeatletter
\providecommand \@ifxundefined [1]{%
 \@ifx{#1\undefined}
}%
\providecommand \@ifnum [1]{%
 \ifnum #1\expandafter \@firstoftwo
 \else \expandafter \@secondoftwo
 \fi
}%
\providecommand \@ifx [1]{%
 \ifx #1\expandafter \@firstoftwo
 \else \expandafter \@secondoftwo
 \fi
}%
\providecommand \natexlab [1]{#1}%
\providecommand \enquote  [1]{``#1''}%
\providecommand \bibnamefont  [1]{#1}%
\providecommand \bibfnamefont [1]{#1}%
\providecommand \citenamefont [1]{#1}%
\providecommand \href@noop [0]{\@secondoftwo}%
\providecommand \href [0]{\begingroup \@sanitize@url \@href}%
\providecommand \@href[1]{\@@startlink{#1}\@@href}%
\providecommand \@@href[1]{\endgroup#1\@@endlink}%
\providecommand \@sanitize@url [0]{\catcode `\\12\catcode `\$12\catcode
  `\&12\catcode `\#12\catcode `\^12\catcode `\_12\catcode `\%12\relax}%
\providecommand \@@startlink[1]{}%
\providecommand \@@endlink[0]{}%
\providecommand \url  [0]{\begingroup\@sanitize@url \@url }%
\providecommand \@url [1]{\endgroup\@href {#1}{\urlprefix }}%
\providecommand \urlprefix  [0]{URL }%
\providecommand \Eprint [0]{\href }%
\providecommand \doibase [0]{http://dx.doi.org/}%
\providecommand \selectlanguage [0]{\@gobble}%
\providecommand \bibinfo  [0]{\@secondoftwo}%
\providecommand \bibfield  [0]{\@secondoftwo}%
\providecommand \translation [1]{[#1]}%
\providecommand \BibitemOpen [0]{}%
\providecommand \bibitemStop [0]{}%
\providecommand \bibitemNoStop [0]{.\EOS\space}%
\providecommand \EOS [0]{\spacefactor3000\relax}%
\providecommand \BibitemShut  [1]{\csname bibitem#1\endcsname}%
\let\auto@bib@innerbib\@empty
\bibitem [{\citenamefont {Mead}(1964)}]{mead}%
  \BibitemOpen
  \bibfield  {author} {\bibinfo {author} {\bibfnamefont {C.~A.}\ \bibnamefont
  {Mead}},\ }\href@noop {} {\bibfield  {journal} {\bibinfo  {journal} {Phys.
  Rev.}\ }\textbf {\bibinfo {volume} {135}},\ \bibinfo {pages} {B849} (\bibinfo
  {year} {1964})}\BibitemShut {NoStop}%
\bibitem [{\citenamefont {Maggiore}(1993{\natexlab{a}})}]{maggiore}%
  \BibitemOpen
  \bibfield  {author} {\bibinfo {author} {\bibfnamefont {M.}~\bibnamefont
  {Maggiore}},\ }\href@noop {} {\bibfield  {journal} {\bibinfo  {journal}
  {Phys. Lett. B}\ }\textbf {\bibinfo {volume} {304}},\ \bibinfo {pages} {65}
  (\bibinfo {year} {1993}{\natexlab{a}})}\BibitemShut {NoStop}%
\bibitem [{\citenamefont {Scardigli}(1999)}]{scardigli}%
  \BibitemOpen
  \bibfield  {author} {\bibinfo {author} {\bibfnamefont {F.}~\bibnamefont
  {Scardigli}},\ }\href@noop {} {\bibfield  {journal} {\bibinfo  {journal}
  {Phys. Lett. B}\ }\textbf {\bibinfo {volume} {452}},\ \bibinfo {pages} {39}
  (\bibinfo {year} {1999})}\BibitemShut {NoStop}%
\bibitem [{\citenamefont {Garay}(1995)}]{garay}%
  \BibitemOpen
  \bibfield  {author} {\bibinfo {author} {\bibfnamefont {L.~J.}\ \bibnamefont
  {Garay}},\ }\href@noop {} {\bibfield  {journal} {\bibinfo  {journal} {Int. J.
  Mod. Phys. A}\ }\textbf {\bibinfo {volume} {10}},\ \bibinfo {pages} {145}
  (\bibinfo {year} {1995})}\BibitemShut {NoStop}%
\bibitem [{\citenamefont {Hossenfelder}(2013)}]{hossen2013}%
  \BibitemOpen
  \bibfield  {author} {\bibinfo {author} {\bibfnamefont {S.}~\bibnamefont
  {Hossenfelder}},\ }\href@noop {} {\bibfield  {journal} {\bibinfo  {journal}
  {Liv. Rev. Relativ.}\ }\textbf {\bibinfo {volume} {16}},\ \bibinfo {pages}
  {2} (\bibinfo {year} {2013})}\BibitemShut {NoStop}%
\bibitem [{\citenamefont {Tawfik}\ and\ \citenamefont
  {Diab}(2014)}]{tawfik-diab-2014}%
  \BibitemOpen
  \bibfield  {author} {\bibinfo {author} {\bibfnamefont {A.}~\bibnamefont
  {Tawfik}}\ and\ \bibinfo {author} {\bibfnamefont {A.}~\bibnamefont {Diab}},\
  }\href@noop {} {\bibfield  {journal} {\bibinfo  {journal} {Int. J. Mod. Phys.
  D}\ }\textbf {\bibinfo {volume} {23}},\ \bibinfo {pages} {1430025} (\bibinfo
  {year} {2014})}\BibitemShut {NoStop}%
\bibitem [{\citenamefont {Tawfik}\ and\ \citenamefont
  {Diab}(2015)}]{tawfik-diab-2015}%
  \BibitemOpen
  \bibfield  {author} {\bibinfo {author} {\bibfnamefont {A.}~\bibnamefont
  {Tawfik}}\ and\ \bibinfo {author} {\bibfnamefont {A.}~\bibnamefont {Diab}},\
  }\href@noop {} {\bibfield  {journal} {\bibinfo  {journal} {Rep. Prog. Phys.}\
  }\textbf {\bibinfo {volume} {78}},\ \bibinfo {pages} {126001} (\bibinfo
  {year} {2015})}\BibitemShut {NoStop}%
\bibitem [{\citenamefont {Gomes}(2023{\natexlab{a}})}]{igup}%
  \BibitemOpen
  \bibfield  {author} {\bibinfo {author} {\bibfnamefont {A.~H.}\ \bibnamefont
  {Gomes}},\ }\href@noop {} {\bibfield  {journal} {\bibinfo  {journal} {J.
  Phys. A: Math. Theor.}\ }\textbf {\bibinfo {volume} {56}},\ \bibinfo {pages}
  {035301} (\bibinfo {year} {2023}{\natexlab{a}})}\BibitemShut {NoStop}%
\bibitem [{\citenamefont {Bosso}\ \emph
  {et~al.}(2023{\natexlab{a}})\citenamefont {Bosso}, \citenamefont {Luciano},
  \citenamefont {Petruzziello},\ and\ \citenamefont
  {Wagner}}]{bosso-luciano-petruzziello-wagner}%
  \BibitemOpen
  \bibfield  {author} {\bibinfo {author} {\bibfnamefont {P.}~\bibnamefont
  {Bosso}}, \bibinfo {author} {\bibfnamefont {G.~G.}\ \bibnamefont {Luciano}},
  \bibinfo {author} {\bibfnamefont {L.}~\bibnamefont {Petruzziello}}, \ and\
  \bibinfo {author} {\bibfnamefont {F.}~\bibnamefont {Wagner}},\ }\href@noop {}
  {\bibfield  {journal} {\bibinfo  {journal} {arXiv:2305.16193v1 [gr-qc]}\ }
  (\bibinfo {year} {2023}{\natexlab{a}})}\BibitemShut {NoStop}%
\bibitem [{\citenamefont {Zhu}\ \emph {et~al.}(2009)\citenamefont {Zhu},
  \citenamefont {Ren},\ and\ \citenamefont {Li}}]{cosmo1}%
  \BibitemOpen
  \bibfield  {author} {\bibinfo {author} {\bibfnamefont {T.}~\bibnamefont
  {Zhu}}, \bibinfo {author} {\bibfnamefont {J.-R.}\ \bibnamefont {Ren}}, \ and\
  \bibinfo {author} {\bibfnamefont {M.-F.}\ \bibnamefont {Li}},\ }\href@noop {}
  {\bibfield  {journal} {\bibinfo  {journal} {Phys. Lett B}\ }\textbf {\bibinfo
  {volume} {674}},\ \bibinfo {pages} {204} (\bibinfo {year}
  {2009})}\BibitemShut {NoStop}%
\bibitem [{\citenamefont {Scardigli}\ and\ \citenamefont
  {Casadio}(2015)}]{scardigli-casadio}%
  \BibitemOpen
  \bibfield  {author} {\bibinfo {author} {\bibfnamefont {F.}~\bibnamefont
  {Scardigli}}\ and\ \bibinfo {author} {\bibfnamefont {R.}~\bibnamefont
  {Casadio}},\ }\href@noop {} {\bibfield  {journal} {\bibinfo  {journal} {Eur.
  Phys. J. C}\ }\textbf {\bibinfo {volume} {75}},\ \bibinfo {pages} {425}
  (\bibinfo {year} {2015})}\BibitemShut {NoStop}%
\bibitem [{\citenamefont {Ong}(2018)}]{grav1}%
  \BibitemOpen
  \bibfield  {author} {\bibinfo {author} {\bibfnamefont {Y.~C.}\ \bibnamefont
  {Ong}},\ }\href@noop {} {\bibfield  {journal} {\bibinfo  {journal} {J.
  Cosmol. Astropart. Phys.}\ }\textbf {\bibinfo {volume} {09}},\ \bibinfo
  {pages} {015} (\bibinfo {year} {2018})}\BibitemShut {NoStop}%
\bibitem [{\citenamefont {Ong}\ and\ \citenamefont {Yao}(2018)}]{grav2}%
  \BibitemOpen
  \bibfield  {author} {\bibinfo {author} {\bibfnamefont {Y.~C.}\ \bibnamefont
  {Ong}}\ and\ \bibinfo {author} {\bibfnamefont {Y.}~\bibnamefont {Yao}},\
  }\href@noop {} {\bibfield  {journal} {\bibinfo  {journal} {Phys. Rev. D}\
  }\textbf {\bibinfo {volume} {98}},\ \bibinfo {pages} {126018} (\bibinfo
  {year} {2018})}\BibitemShut {NoStop}%
\bibitem [{\citenamefont {Buoninfante}\ \emph {et~al.}(2020)\citenamefont
  {Buoninfante}, \citenamefont {Lambiase}, \citenamefont {Luciano},\ and\
  \citenamefont {Petruzziello}}]{stars}%
  \BibitemOpen
  \bibfield  {author} {\bibinfo {author} {\bibfnamefont {L.}~\bibnamefont
  {Buoninfante}}, \bibinfo {author} {\bibfnamefont {G.}~\bibnamefont
  {Lambiase}}, \bibinfo {author} {\bibfnamefont {G.~G.}\ \bibnamefont
  {Luciano}}, \ and\ \bibinfo {author} {\bibfnamefont {L.}~\bibnamefont
  {Petruzziello}},\ }\href@noop {} {\bibfield  {journal} {\bibinfo  {journal}
  {Eur. Phys. J. C}\ }\textbf {\bibinfo {volume} {80}},\ \bibinfo {pages} {853}
  (\bibinfo {year} {2020})}\BibitemShut {NoStop}%
\bibitem [{\citenamefont {Casadio}\ and\ \citenamefont
  {Scardigli}(2020)}]{casadio-scardigli-2020}%
  \BibitemOpen
  \bibfield  {author} {\bibinfo {author} {\bibfnamefont {R.}~\bibnamefont
  {Casadio}}\ and\ \bibinfo {author} {\bibfnamefont {F.}~\bibnamefont
  {Scardigli}},\ }\href@noop {} {\bibfield  {journal} {\bibinfo  {journal}
  {Phys. Lett. B}\ }\textbf {\bibinfo {volume} {807}},\ \bibinfo {pages}
  {135558} (\bibinfo {year} {2020})}\BibitemShut {NoStop}%
\bibitem [{\citenamefont {Giardino}\ and\ \citenamefont
  {Salzano}(2021)}]{cosmo2}%
  \BibitemOpen
  \bibfield  {author} {\bibinfo {author} {\bibfnamefont {S.}~\bibnamefont
  {Giardino}}\ and\ \bibinfo {author} {\bibfnamefont {V.}~\bibnamefont
  {Salzano}},\ }\href@noop {} {\bibfield  {journal} {\bibinfo  {journal} {Eur.
  Phys. J. C}\ }\textbf {\bibinfo {volume} {81}},\ \bibinfo {pages} {110}
  (\bibinfo {year} {2021})}\BibitemShut {NoStop}%
\bibitem [{\citenamefont {Gregoris}\ and\ \citenamefont {Ong}(2023)}]{grav3}%
  \BibitemOpen
  \bibfield  {author} {\bibinfo {author} {\bibfnamefont {D.}~\bibnamefont
  {Gregoris}}\ and\ \bibinfo {author} {\bibfnamefont {Y.~C.}\ \bibnamefont
  {Ong}},\ }\href@noop {} {\bibfield  {journal} {\bibinfo  {journal} {Ann.
  Phys.}\ }\textbf {\bibinfo {volume} {452}},\ \bibinfo {pages} {169287}
  (\bibinfo {year} {2023})}\BibitemShut {NoStop}%
\bibitem [{\citenamefont {Maggiore}(1993{\natexlab{b}})}]{maggiore-algebra}%
  \BibitemOpen
  \bibfield  {author} {\bibinfo {author} {\bibfnamefont {M.}~\bibnamefont
  {Maggiore}},\ }\href@noop {} {\bibfield  {journal} {\bibinfo  {journal}
  {Phys. Lett. B}\ }\textbf {\bibinfo {volume} {319}},\ \bibinfo {pages} {83}
  (\bibinfo {year} {1993}{\natexlab{b}})}\BibitemShut {NoStop}%
\bibitem [{\citenamefont {Kempf}\ \emph {et~al.}(1995)\citenamefont {Kempf},
  \citenamefont {Mangano},\ and\ \citenamefont {Mann}}]{kmm95}%
  \BibitemOpen
  \bibfield  {author} {\bibinfo {author} {\bibfnamefont {A.}~\bibnamefont
  {Kempf}}, \bibinfo {author} {\bibfnamefont {G.}~\bibnamefont {Mangano}}, \
  and\ \bibinfo {author} {\bibfnamefont {R.~B.}\ \bibnamefont {Mann}},\
  }\href@noop {} {\bibfield  {journal} {\bibinfo  {journal} {Phys. Rev. D}\
  }\textbf {\bibinfo {volume} {52}},\ \bibinfo {pages} {1108} (\bibinfo {year}
  {1995})}\BibitemShut {NoStop}%
\bibitem [{\citenamefont {Robertson}(1929)}]{robertson}%
  \BibitemOpen
  \bibfield  {author} {\bibinfo {author} {\bibfnamefont {H.~P.}\ \bibnamefont
  {Robertson}},\ }\href@noop {} {\bibfield  {journal} {\bibinfo  {journal}
  {Phys. Rev.}\ }\textbf {\bibinfo {volume} {34}},\ \bibinfo {pages} {163}
  (\bibinfo {year} {1929})}\BibitemShut {NoStop}%
\bibitem [{\citenamefont {Schr\"odinger}(1930)}]{schrodinger}%
  \BibitemOpen
  \bibfield  {author} {\bibinfo {author} {\bibfnamefont {E.}~\bibnamefont
  {Schr\"odinger}},\ }\href@noop {} {\bibfield  {journal} {\bibinfo  {journal}
  {Sitzungsberichte der Preuss. Akad. der Wissenschaften. Phys. Klasse}\
  }\textbf {\bibinfo {volume} {14}},\ \bibinfo {pages} {296} (\bibinfo {year}
  {1930})}\BibitemShut {NoStop}%
\bibitem [{\citenamefont {Abdelkhalek}\ \emph {et~al.}(2016)\citenamefont
  {Abdelkhalek}, \citenamefont {Chemissany}, \citenamefont {Fiedler},
  \citenamefont {Mangano},\ and\ \citenamefont {Schwonnek}}]{mangano2016}%
  \BibitemOpen
  \bibfield  {author} {\bibinfo {author} {\bibfnamefont {K.}~\bibnamefont
  {Abdelkhalek}}, \bibinfo {author} {\bibfnamefont {W.}~\bibnamefont
  {Chemissany}}, \bibinfo {author} {\bibfnamefont {L.}~\bibnamefont {Fiedler}},
  \bibinfo {author} {\bibfnamefont {G.}~\bibnamefont {Mangano}}, \ and\
  \bibinfo {author} {\bibfnamefont {R.}~\bibnamefont {Schwonnek}},\ }\href@noop
  {} {\bibfield  {journal} {\bibinfo  {journal} {Phys. Rev. D}\ }\textbf
  {\bibinfo {volume} {94}},\ \bibinfo {pages} {123505} (\bibinfo {year}
  {2016})}\BibitemShut {NoStop}%
\bibitem [{\citenamefont {Bosso}\ \emph
  {et~al.}(2023{\natexlab{b}})\citenamefont {Bosso}, \citenamefont
  {Petruzziello},\ and\ \citenamefont {Wagner}}]{bosso-petruzziello-wagner}%
  \BibitemOpen
  \bibfield  {author} {\bibinfo {author} {\bibfnamefont {P.}~\bibnamefont
  {Bosso}}, \bibinfo {author} {\bibfnamefont {L.}~\bibnamefont {Petruzziello}},
  \ and\ \bibinfo {author} {\bibfnamefont {F.}~\bibnamefont {Wagner}},\
  }\href@noop {} {\bibfield  {journal} {\bibinfo  {journal} {arXiv:2302.04564v1
  [hep-th]}\ } (\bibinfo {year} {2023}{\natexlab{b}})}\BibitemShut {NoStop}%
\bibitem [{\citenamefont {Detournay}\ \emph {et~al.}(2002)\citenamefont
  {Detournay}, \citenamefont {Gabriel},\ and\ \citenamefont
  {Spindel}}]{detournay}%
  \BibitemOpen
  \bibfield  {author} {\bibinfo {author} {\bibfnamefont {S.}~\bibnamefont
  {Detournay}}, \bibinfo {author} {\bibfnamefont {C.}~\bibnamefont {Gabriel}},
  \ and\ \bibinfo {author} {\bibfnamefont {P.}~\bibnamefont {Spindel}},\
  }\href@noop {} {\bibfield  {journal} {\bibinfo  {journal} {Phys. Rev. D}\
  }\textbf {\bibinfo {volume} {66}},\ \bibinfo {pages} {125004} (\bibinfo
  {year} {2002})}\BibitemShut {NoStop}%
\bibitem [{\citenamefont {Bernardo}\ and\ \citenamefont
  {Esguerra}(2018)}]{bernardo-esguerra-maximally2}%
  \BibitemOpen
  \bibfield  {author} {\bibinfo {author} {\bibfnamefont {R.~C.~S.}\
  \bibnamefont {Bernardo}}\ and\ \bibinfo {author} {\bibfnamefont {J.~P.~H.}\
  \bibnamefont {Esguerra}},\ }\href@noop {} {\bibfield  {journal} {\bibinfo
  {journal} {Ann. Phys.}\ }\textbf {\bibinfo {volume} {391}},\ \bibinfo {pages}
  {293} (\bibinfo {year} {2018})}\BibitemShut {NoStop}%
\bibitem [{\citenamefont {Bosso}(2021)}]{pasquale-xbasis}%
  \BibitemOpen
  \bibfield  {author} {\bibinfo {author} {\bibfnamefont {P.}~\bibnamefont
  {Bosso}},\ }\href@noop {} {\bibfield  {journal} {\bibinfo  {journal} {Class.
  Quantum Grav.}\ }\textbf {\bibinfo {volume} {38}},\ \bibinfo {pages} {075021}
  (\bibinfo {year} {2021})}\BibitemShut {NoStop}%
\bibitem [{\citenamefont {Nozari}\ and\ \citenamefont
  {Azizi}(2006)}]{nozari2006}%
  \BibitemOpen
  \bibfield  {author} {\bibinfo {author} {\bibfnamefont {K.}~\bibnamefont
  {Nozari}}\ and\ \bibinfo {author} {\bibfnamefont {T.}~\bibnamefont {Azizi}},\
  }\href@noop {} {\bibfield  {journal} {\bibinfo  {journal} {Gen. Relativ.
  Gravit.}\ }\textbf {\bibinfo {volume} {38}},\ \bibinfo {pages} {735}
  (\bibinfo {year} {2006})}\BibitemShut {NoStop}%
\bibitem [{\citenamefont {Ali}\ \emph {et~al.}(2009)\citenamefont {Ali},
  \citenamefont {Das},\ and\ \citenamefont {Vagenas}}]{ali-das-vagenas09}%
  \BibitemOpen
  \bibfield  {author} {\bibinfo {author} {\bibfnamefont {A.~F.}\ \bibnamefont
  {Ali}}, \bibinfo {author} {\bibfnamefont {S.}~\bibnamefont {Das}}, \ and\
  \bibinfo {author} {\bibfnamefont {E.~C.}\ \bibnamefont {Vagenas}},\
  }\href@noop {} {\bibfield  {journal} {\bibinfo  {journal} {Phys. Lett. B}\
  }\textbf {\bibinfo {volume} {678}},\ \bibinfo {pages} {497} (\bibinfo {year}
  {2009})}\BibitemShut {NoStop}%
\bibitem [{\citenamefont {Pedram}(2010{\natexlab{a}})}]{pedram-europhys}%
  \BibitemOpen
  \bibfield  {author} {\bibinfo {author} {\bibfnamefont {P.}~\bibnamefont
  {Pedram}},\ }\href@noop {} {\bibfield  {journal} {\bibinfo  {journal}
  {Europhys. Lett.}\ }\textbf {\bibinfo {volume} {89}},\ \bibinfo {pages}
  {50008} (\bibinfo {year} {2010}{\natexlab{a}})}\BibitemShut {NoStop}%
\bibitem [{\citenamefont {Pedram}(2010{\natexlab{b}})}]{pedram2010}%
  \BibitemOpen
  \bibfield  {author} {\bibinfo {author} {\bibfnamefont {P.}~\bibnamefont
  {Pedram}},\ }\href@noop {} {\bibfield  {journal} {\bibinfo  {journal} {Int.
  J. Mod. Phys. D}\ }\textbf {\bibinfo {volume} {19}},\ \bibinfo {pages} {2003}
  (\bibinfo {year} {2010}{\natexlab{b}})}\BibitemShut {NoStop}%
\bibitem [{\citenamefont {Ali}\ \emph {et~al.}(2011)\citenamefont {Ali},
  \citenamefont {Das},\ and\ \citenamefont {Vagenas}}]{ali-das-vagenas11}%
  \BibitemOpen
  \bibfield  {author} {\bibinfo {author} {\bibfnamefont {A.~F.}\ \bibnamefont
  {Ali}}, \bibinfo {author} {\bibfnamefont {S.}~\bibnamefont {Das}}, \ and\
  \bibinfo {author} {\bibfnamefont {E.~C.}\ \bibnamefont {Vagenas}},\
  }\href@noop {} {\bibfield  {journal} {\bibinfo  {journal} {Phys. Rev. D}\
  }\textbf {\bibinfo {volume} {84}},\ \bibinfo {pages} {044013} (\bibinfo
  {year} {2011})}\BibitemShut {NoStop}%
\bibitem [{\citenamefont {Pedram}(2012{\natexlab{a}})}]{pedram12-plb3}%
  \BibitemOpen
  \bibfield  {author} {\bibinfo {author} {\bibfnamefont {P.}~\bibnamefont
  {Pedram}},\ }\href@noop {} {\bibfield  {journal} {\bibinfo  {journal} {Phys.
  Lett. B}\ }\textbf {\bibinfo {volume} {718}},\ \bibinfo {pages} {638}
  (\bibinfo {year} {2012}{\natexlab{a}})}\BibitemShut {NoStop}%
\bibitem [{\citenamefont {Pedram}(2012{\natexlab{b}})}]{pedram12-prd}%
  \BibitemOpen
  \bibfield  {author} {\bibinfo {author} {\bibfnamefont {P.}~\bibnamefont
  {Pedram}},\ }\href@noop {} {\bibfield  {journal} {\bibinfo  {journal} {Phys.
  Rev. D}\ }\textbf {\bibinfo {volume} {85}},\ \bibinfo {pages} {024016}
  (\bibinfo {year} {2012}{\natexlab{b}})}\BibitemShut {NoStop}%
\bibitem [{\citenamefont {Blado}\ \emph {et~al.}(2014)\citenamefont {Blado},
  \citenamefont {Owens},\ and\ \citenamefont {Meyers}}]{blado}%
  \BibitemOpen
  \bibfield  {author} {\bibinfo {author} {\bibfnamefont {G.}~\bibnamefont
  {Blado}}, \bibinfo {author} {\bibfnamefont {C.}~\bibnamefont {Owens}}, \ and\
  \bibinfo {author} {\bibfnamefont {V.}~\bibnamefont {Meyers}},\ }\href@noop {}
  {\bibfield  {journal} {\bibinfo  {journal} {Eur. J. Phys.}\ }\textbf
  {\bibinfo {volume} {35}},\ \bibinfo {pages} {065011} (\bibinfo {year}
  {2014})}\BibitemShut {NoStop}%
\bibitem [{\citenamefont {Bernardo}\ and\ \citenamefont
  {Esguerra}(2016{\natexlab{a}})}]{bernardo-esguerra}%
  \BibitemOpen
  \bibfield  {author} {\bibinfo {author} {\bibfnamefont {R.~C.~S.}\
  \bibnamefont {Bernardo}}\ and\ \bibinfo {author} {\bibfnamefont {J.~P.~H.}\
  \bibnamefont {Esguerra}},\ }\href@noop {} {\bibfield  {journal} {\bibinfo
  {journal} {Ann. Phys.}\ }\textbf {\bibinfo {volume} {373}},\ \bibinfo {pages}
  {521} (\bibinfo {year} {2016}{\natexlab{a}})}\BibitemShut {NoStop}%
\bibitem [{\citenamefont {Shababi}\ \emph {et~al.}(2016)\citenamefont
  {Shababi}, \citenamefont {Pedram},\ and\ \citenamefont {Chung}}]{Shababi16}%
  \BibitemOpen
  \bibfield  {author} {\bibinfo {author} {\bibfnamefont {H.}~\bibnamefont
  {Shababi}}, \bibinfo {author} {\bibfnamefont {P.}~\bibnamefont {Pedram}}, \
  and\ \bibinfo {author} {\bibfnamefont {W.~S.}\ \bibnamefont {Chung}},\
  }\href@noop {} {\bibfield  {journal} {\bibinfo  {journal} {Int. J. Mod. Phys.
  A}\ }\textbf {\bibinfo {volume} {31}},\ \bibinfo {pages} {1650101} (\bibinfo
  {year} {2016})}\BibitemShut {NoStop}%
\bibitem [{\citenamefont {Shababi}\ and\ \citenamefont
  {Chung}(2018)}]{Shababi18}%
  \BibitemOpen
  \bibfield  {author} {\bibinfo {author} {\bibfnamefont {H.}~\bibnamefont
  {Shababi}}\ and\ \bibinfo {author} {\bibfnamefont {W.~S.}\ \bibnamefont
  {Chung}},\ }\href@noop {} {\bibfield  {journal} {\bibinfo  {journal} {Mod.
  Phys. Lett. A}\ }\textbf {\bibinfo {volume} {33}},\ \bibinfo {pages}
  {1850068} (\bibinfo {year} {2018})}\BibitemShut {NoStop}%
\bibitem [{\citenamefont {Chung}\ and\ \citenamefont
  {Hassanabadi}(2019)}]{chung-hass2019}%
  \BibitemOpen
  \bibfield  {author} {\bibinfo {author} {\bibfnamefont {W.~S.}\ \bibnamefont
  {Chung}}\ and\ \bibinfo {author} {\bibfnamefont {H.}~\bibnamefont
  {Hassanabadi}},\ }\href@noop {} {\bibfield  {journal} {\bibinfo  {journal}
  {Eur. Phys. J. C}\ }\textbf {\bibinfo {volume} {79}},\ \bibinfo {pages} {213}
  (\bibinfo {year} {2019})}\BibitemShut {NoStop}%
\bibitem [{\citenamefont {A.~O.~O}\ \emph {et~al.}(2020)\citenamefont
  {A.~O.~O}, \citenamefont {Gusson}, \citenamefont {Dilem}, \citenamefont
  {Furtado}, \citenamefont {Francisco}, \citenamefont {Fabris},\ and\
  \citenamefont {Nogueira}}]{nogueira2020}%
  \BibitemOpen
  \bibfield  {author} {\bibinfo {author} {\bibfnamefont {G.}~\bibnamefont
  {A.~O.~O}}, \bibinfo {author} {\bibfnamefont {M.~F.}\ \bibnamefont {Gusson}},
  \bibinfo {author} {\bibfnamefont {B.~B.}\ \bibnamefont {Dilem}}, \bibinfo
  {author} {\bibfnamefont {R.~G.}\ \bibnamefont {Furtado}}, \bibinfo {author}
  {\bibfnamefont {R.~O.}\ \bibnamefont {Francisco}}, \bibinfo {author}
  {\bibfnamefont {J.~C.}\ \bibnamefont {Fabris}}, \ and\ \bibinfo {author}
  {\bibfnamefont {J.~A.}\ \bibnamefont {Nogueira}},\ }\href@noop {} {\bibfield
  {journal} {\bibinfo  {journal} {Int. J. Mod. Phys. A}\ }\textbf {\bibinfo
  {volume} {35}},\ \bibinfo {pages} {2050069} (\bibinfo {year}
  {2020})}\BibitemShut {NoStop}%
\bibitem [{\citenamefont {Shababi}\ and\ \citenamefont
  {Chung}(2020)}]{Shababi20}%
  \BibitemOpen
  \bibfield  {author} {\bibinfo {author} {\bibfnamefont {H.}~\bibnamefont
  {Shababi}}\ and\ \bibinfo {author} {\bibfnamefont {W.~S.}\ \bibnamefont
  {Chung}},\ }\href@noop {} {\bibfield  {journal} {\bibinfo  {journal} {Mod.
  Phys. Lett. A}\ }\textbf {\bibinfo {volume} {35}},\ \bibinfo {pages}
  {2050018} (\bibinfo {year} {2020})}\BibitemShut {NoStop}%
\bibitem [{\citenamefont {Bernardo}\ and\ \citenamefont
  {Esguerra}(2016{\natexlab{b}})}]{scattering}%
  \BibitemOpen
  \bibfield  {author} {\bibinfo {author} {\bibfnamefont {R.~C.~S.}\
  \bibnamefont {Bernardo}}\ and\ \bibinfo {author} {\bibfnamefont {J.~P.~H.}\
  \bibnamefont {Esguerra}},\ }\href@noop {} {\bibfield  {journal} {\bibinfo
  {journal} {Ann. Phys.}\ }\textbf {\bibinfo {volume} {375}},\ \bibinfo {pages}
  {444} (\bibinfo {year} {2016}{\natexlab{b}})}\BibitemShut {NoStop}%
\bibitem [{\citenamefont {Bosso}\ and\ \citenamefont
  {Luciano}(2021)}]{bosso-luciano}%
  \BibitemOpen
  \bibfield  {author} {\bibinfo {author} {\bibfnamefont {P.}~\bibnamefont
  {Bosso}}\ and\ \bibinfo {author} {\bibfnamefont {G.~G.}\ \bibnamefont
  {Luciano}},\ }\href@noop {} {\bibfield  {journal} {\bibinfo  {journal} {Eur.
  Phys. J. C}\ }\textbf {\bibinfo {volume} {81}},\ \bibinfo {pages} {982}
  (\bibinfo {year} {2021})}\BibitemShut {NoStop}%
\bibitem [{\citenamefont {Bernardo}\ and\ \citenamefont
  {Esguerra}(2017)}]{bernardo-esguerra-maximally1}%
  \BibitemOpen
  \bibfield  {author} {\bibinfo {author} {\bibfnamefont {R.~C.~S.}\
  \bibnamefont {Bernardo}}\ and\ \bibinfo {author} {\bibfnamefont {J.~P.~H.}\
  \bibnamefont {Esguerra}},\ }\href@noop {} {\bibfield  {journal} {\bibinfo
  {journal} {J. Math. Phys.}\ }\textbf {\bibinfo {volume} {58}},\ \bibinfo
  {pages} {042103} (\bibinfo {year} {2017})}\BibitemShut {NoStop}%
\bibitem [{\citenamefont {Lubo}(2002)}]{lubo}%
  \BibitemOpen
  \bibfield  {author} {\bibinfo {author} {\bibfnamefont {M.}~\bibnamefont
  {Lubo}},\ }\href@noop {} {\bibfield  {journal} {\bibinfo  {journal} {Phys.
  Rev. D}\ }\textbf {\bibinfo {volume} {55}},\ \bibinfo {pages} {066003}
  (\bibinfo {year} {2002})}\BibitemShut {NoStop}%
\bibitem [{\citenamefont {Frassino}\ and\ \citenamefont
  {Panella}(2012)}]{casimir1}%
  \BibitemOpen
  \bibfield  {author} {\bibinfo {author} {\bibfnamefont {A.~M.}\ \bibnamefont
  {Frassino}}\ and\ \bibinfo {author} {\bibfnamefont {O.}~\bibnamefont
  {Panella}},\ }\href@noop {} {\bibfield  {journal} {\bibinfo  {journal} {Phys.
  Rev. D}\ }\textbf {\bibinfo {volume} {85}},\ \bibinfo {pages} {045030}
  (\bibinfo {year} {2012})}\BibitemShut {NoStop}%
\bibitem [{\citenamefont {Dorsch}\ and\ \citenamefont
  {Nogueira}(2012)}]{casimir2}%
  \BibitemOpen
  \bibfield  {author} {\bibinfo {author} {\bibfnamefont {G.~C.}\ \bibnamefont
  {Dorsch}}\ and\ \bibinfo {author} {\bibfnamefont {J.~A.}\ \bibnamefont
  {Nogueira}},\ }\href@noop {} {\bibfield  {journal} {\bibinfo  {journal} {Int.
  J. Mod. Phys. A}\ }\textbf {\bibinfo {volume} {27}},\ \bibinfo {pages}
  {1250113} (\bibinfo {year} {2012})}\BibitemShut {NoStop}%
\bibitem [{\citenamefont {Blasone}\ \emph {et~al.}(2020)\citenamefont
  {Blasone}, \citenamefont {Lambiase}, \citenamefont {Luciano}, \citenamefont
  {Petruzziello},\ and\ \citenamefont {Scardigli}}]{casimir3}%
  \BibitemOpen
  \bibfield  {author} {\bibinfo {author} {\bibfnamefont {M.}~\bibnamefont
  {Blasone}}, \bibinfo {author} {\bibfnamefont {G.}~\bibnamefont {Lambiase}},
  \bibinfo {author} {\bibfnamefont {G.~G.}\ \bibnamefont {Luciano}}, \bibinfo
  {author} {\bibfnamefont {L.}~\bibnamefont {Petruzziello}}, \ and\ \bibinfo
  {author} {\bibfnamefont {F.}~\bibnamefont {Scardigli}},\ }\href@noop {}
  {\bibfield  {journal} {\bibinfo  {journal} {Int. J. Mod. Phys. D}\ }\textbf
  {\bibinfo {volume} {29}},\ \bibinfo {pages} {2050011} (\bibinfo {year}
  {2020})}\BibitemShut {NoStop}%
\bibitem [{\citenamefont {Gomes}(2023{\natexlab{b}})}]{andre-canonical}%
  \BibitemOpen
  \bibfield  {author} {\bibinfo {author} {\bibfnamefont {A.~H.}\ \bibnamefont
  {Gomes}},\ }\href@noop {} {\bibfield  {journal} {\bibinfo  {journal} {Class.
  Quantum Grav.}\ }\textbf {\bibinfo {volume} {40}},\ \bibinfo {pages} {065005}
  (\bibinfo {year} {2023}{\natexlab{b}})}\BibitemShut {NoStop}%
\bibitem [{\citenamefont {Kempf}\ and\ \citenamefont
  {Mangano}(1997)}]{kempf-mangano97}%
  \BibitemOpen
  \bibfield  {author} {\bibinfo {author} {\bibfnamefont {A.}~\bibnamefont
  {Kempf}}\ and\ \bibinfo {author} {\bibfnamefont {G.}~\bibnamefont
  {Mangano}},\ }\href@noop {} {\bibfield  {journal} {\bibinfo  {journal} {Phys.
  Rev. D}\ }\textbf {\bibinfo {volume} {55}},\ \bibinfo {pages} {7909}
  (\bibinfo {year} {1997})}\BibitemShut {NoStop}%
\bibitem [{\citenamefont {Cohen-Tannoudji}\ \emph {et~al.}(1977)\citenamefont
  {Cohen-Tannoudji}, \citenamefont {Diu},\ and\ \citenamefont
  {Lalo\"e}}]{cohen1}%
  \BibitemOpen
  \bibfield  {author} {\bibinfo {author} {\bibfnamefont {C.}~\bibnamefont
  {Cohen-Tannoudji}}, \bibinfo {author} {\bibfnamefont {B.}~\bibnamefont
  {Diu}}, \ and\ \bibinfo {author} {\bibfnamefont {F.}~\bibnamefont
  {Lalo\"e}},\ }\href@noop {} {\emph {\bibinfo {title} {Quantum Mechanics, Vol.
  1}}}\ (\bibinfo  {publisher} {John Wiley \& Sons},\ \bibinfo {address} {New
  York},\ \bibinfo {year} {1977})\BibitemShut {NoStop}%
\bibitem [{\citenamefont {Kempf}(2000)}]{kempf2000}%
  \BibitemOpen
  \bibfield  {author} {\bibinfo {author} {\bibfnamefont {A.}~\bibnamefont
  {Kempf}},\ }\href@noop {} {\bibfield  {journal} {\bibinfo  {journal} {Phys.
  Rev. D}\ }\textbf {\bibinfo {volume} {63}},\ \bibinfo {pages} {024017}
  (\bibinfo {year} {2000})}\BibitemShut {NoStop}%
\bibitem [{\citenamefont {Pedram}(2012{\natexlab{c}})}]{pedram12-plb2}%
  \BibitemOpen
  \bibfield  {author} {\bibinfo {author} {\bibfnamefont {P.}~\bibnamefont
  {Pedram}},\ }\href@noop {} {\bibfield  {journal} {\bibinfo  {journal} {Phys.
  Lett. B}\ }\textbf {\bibinfo {volume} {714}},\ \bibinfo {pages} {317}
  (\bibinfo {year} {2012}{\natexlab{c}})}\BibitemShut {NoStop}%
\bibitem [{\citenamefont {Nozari}\ and\ \citenamefont
  {Etemadi}(2012)}]{nozari-etemadi}%
  \BibitemOpen
  \bibfield  {author} {\bibinfo {author} {\bibfnamefont {K.}~\bibnamefont
  {Nozari}}\ and\ \bibinfo {author} {\bibfnamefont {A.}~\bibnamefont
  {Etemadi}},\ }\href@noop {} {\bibfield  {journal} {\bibinfo  {journal} {Phys.
  Rev. D}\ }\textbf {\bibinfo {volume} {85}},\ \bibinfo {pages} {104029}
  (\bibinfo {year} {2012})}\BibitemShut {NoStop}%
\bibitem [{\citenamefont {Matsuo}\ and\ \citenamefont
  {Shibusa}(2006)}]{matsuo-shibusa}%
  \BibitemOpen
  \bibfield  {author} {\bibinfo {author} {\bibfnamefont {T.}~\bibnamefont
  {Matsuo}}\ and\ \bibinfo {author} {\bibfnamefont {Y.}~\bibnamefont
  {Shibusa}},\ }\href@noop {} {\bibfield  {journal} {\bibinfo  {journal} {Mod.
  Phys. Lett. A}\ }\textbf {\bibinfo {volume} {21}},\ \bibinfo {pages} {1285}
  (\bibinfo {year} {2006})}\BibitemShut {NoStop}%
\bibitem [{\citenamefont {Das}\ and\ \citenamefont
  {Vagenas}(2008)}]{das-vagenas08}%
  \BibitemOpen
  \bibfield  {author} {\bibinfo {author} {\bibfnamefont {S.}~\bibnamefont
  {Das}}\ and\ \bibinfo {author} {\bibfnamefont {E.~C.}\ \bibnamefont
  {Vagenas}},\ }\href@noop {} {\bibfield  {journal} {\bibinfo  {journal} {Phys.
  Rev. Lett.}\ }\textbf {\bibinfo {volume} {101}},\ \bibinfo {pages} {221301}
  (\bibinfo {year} {2008})}\BibitemShut {NoStop}%
\bibitem [{\citenamefont {Brau}(1999)}]{brau}%
  \BibitemOpen
  \bibfield  {author} {\bibinfo {author} {\bibfnamefont {F.}~\bibnamefont
  {Brau}},\ }\href@noop {} {\bibfield  {journal} {\bibinfo  {journal} {J. Phys.
  A: Math. Gen.}\ }\textbf {\bibinfo {volume} {32}},\ \bibinfo {pages} {7691}
  (\bibinfo {year} {1999})}\BibitemShut {NoStop}%
\end{thebibliography}%

\end{document}